\documentclass[twoside,twocolumn,9pt]{article}
\usepackage{extsizes}
\usepackage[super,sort&compress,comma]{natbib} 
\usepackage[version=3]{mhchem}
\usepackage[left=1.5cm, right=1.5cm, top=1.785cm, bottom=2.0cm]{geometry}
\usepackage{balance}
\usepackage{mathptmx}
\usepackage{sectsty}
\usepackage{graphicx} 
\usepackage{lastpage}
\usepackage[format=plain,justification=justified,singlelinecheck=false,font={stretch=1.125,small,sf},labelfont=bf,labelsep=space]{caption}
\usepackage{float}
\usepackage{fancyhdr}
\usepackage{fnpos}
\usepackage[english]{babel}
\addto{\captionsenglish}{%
  
}
\usepackage{array}
\usepackage{droidsans}
\usepackage{charter}
\usepackage[T1]{fontenc}
\usepackage[usenames,dvipsnames]{xcolor}
\usepackage{setspace}
\usepackage[compact]{titlesec}
\usepackage{hyperref}

\usepackage{epstopdf}

\definecolor{cream}{RGB}{222,217,201}

\begin{document}

\pagestyle{fancy}
\thispagestyle{plain}
\fancypagestyle{plain}{
\renewcommand{\headrulewidth}{0pt}
}

\makeFNbottom
\makeatletter
\renewcommand\LARGE{\@setfontsize\LARGE{15pt}{17}}
\renewcommand\Large{\@setfontsize\Large{12pt}{14}}
\renewcommand\large{\@setfontsize\large{10pt}{12}}
\renewcommand\footnotesize{\@setfontsize\footnotesize{7pt}{10}}
\makeatother

\renewcommand{\thefootnote}{\fnsymbol{footnote}}
\renewcommand\footnoterule{\vspace*{1pt}%
\color{cream}\hrule width 3.5in height 0.4pt \color{black}\vspace*{5pt}} 
\setcounter{secnumdepth}{5}

\makeatletter 
\renewcommand\@biblabel[1]{#1}            
\renewcommand\@makefntext[1]%
{\noindent\makebox[0pt][r]{\@thefnmark\,}#1}
\makeatother 
\renewcommand{\figurename}{\small{Fig.}~}
\sectionfont{\sffamily\Large}
\subsectionfont{\normalsize}
\subsubsectionfont{\bf}
\setstretch{1.125} 
\setlength{\skip\footins}{0.8cm}
\setlength{\footnotesep}{0.25cm}
\setlength{\jot}{10pt}
\titlespacing*{\section}{0pt}{4pt}{4pt}
\titlespacing*{\subsection}{0pt}{15pt}{1pt}

\fancyfoot{}
\fancyfoot[LO,RE]{\vspace{-7.1pt}\includegraphics[height=9pt]{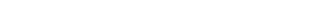}}
\fancyfoot[CO]{\vspace{-7.1pt}\hspace{13.2cm}\includegraphics{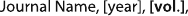}}
\fancyfoot[CE]{\vspace{-7.2pt}\hspace{-14.2cm}\includegraphics{head_foot/RF}}
\fancyfoot[RO]{\footnotesize{\sffamily{1--\pageref{LastPage} ~\textbar  \hspace{2pt}\thepage}}}
\fancyfoot[LE]{\footnotesize{\sffamily{\thepage~\textbar\hspace{3.45cm} 1--\pageref{LastPage}}}}
\fancyhead{}
\renewcommand{\headrulewidth}{0pt} 
\renewcommand{\footrulewidth}{0pt}
\setlength{\arrayrulewidth}{1pt}
\setlength{\columnsep}{6.5mm}
\setlength\bibsep{1pt}

\makeatletter 
\newlength{\figrulesep} 
\setlength{\figrulesep}{0.5\textfloatsep} 

\newcommand{\topfigrule}{\vspace*{-1pt}%
\noindent{\color{cream}\rule[-\figrulesep]{\columnwidth}{1.5pt}} }

\newcommand{\botfigrule}{\vspace*{-2pt}%
\noindent{\color{cream}\rule[\figrulesep]{\columnwidth}{1.5pt}} }

\newcommand{\dblfigrule}{\vspace*{-1pt}%
\noindent{\color{cream}\rule[-\figrulesep]{\textwidth}{1.5pt}} }

\makeatother

\twocolumn[
  \begin{@twocolumnfalse}
{\includegraphics[height=30pt]{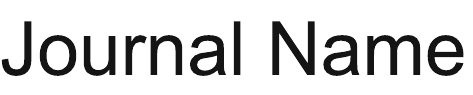}\hfill\raisebox{0pt}[0pt][0pt]{\includegraphics[height=55pt]{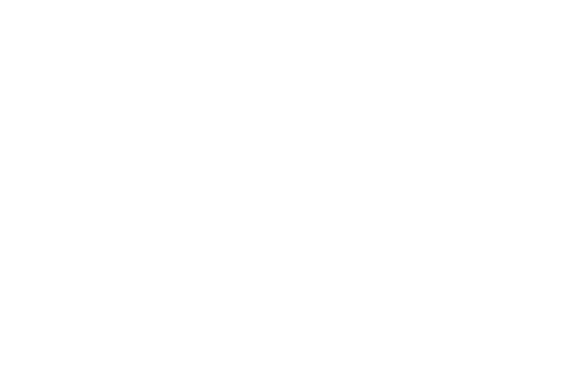}}\\[1ex]
\includegraphics[width=18.5cm]{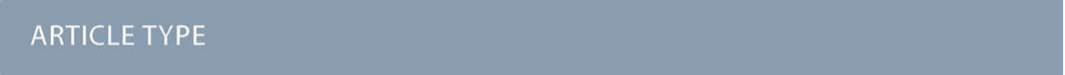}}\par
\vspace{1em}
\sffamily
\begin{tabular}{m{4.5cm} p{13.5cm} }

\includegraphics{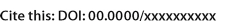} & \noindent\LARGE{\textbf{Particle Image Velocimetry of 3D printed vascular fluidic phantom devices$^\dag$}} \\
\vspace{0.3cm} & \vspace{0.3cm} \\


& \noindent\large{Job van Essen,\textit{$^{a}$} Ahmed Sharaf,\textit{$^{a}$} Denzel Hopman,\textit{$^{a}$} Selene Pirola,\textit{$^{a}$}$^\dag$ and Paola Fanzio\textit{$^{a}$}$^{\ast}$$^\dag$} \\

\includegraphics{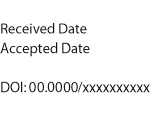} & \noindent\normalsize{Altered hemodynamics play a key role in cerebrovascular diseases such as aneurysms and stenosis. However, in vivo imaging lacks the spatial resolution required to resolve flow dynamics in small vessels.
This study presents an experimental framework to investigate microscale hemodynamics using transparent 3D-printed vascular models and particle image velocimetry (PIV). Optically transparent microfluidic models with straight and pathological (aneurrysmal and stenotic) geometries were fabricated via additive manufacturing up to a minimum diameter size of $500 \mu m$ and characterized using optical microscopy. Flow experiments were conducted under steady laminar conditions, and local velocity fields and wall shear stress (WSS) were measured using micro-PIV.
Measured velocities have been compared with analytical Hagen–Poiseuille predictions, obtaining mean relative errors of $5 - 17 \%$. The platform reliably captured key flow features and spatial variations in velocity. Overall, the results demonstrate that transparent 3D-printed vascular models combined with micro-PIV provide a robust experimental approach for studying microscale cerebrovascular hemodynamics.}\\

\end{tabular}

 \end{@twocolumnfalse} \vspace{0.6cm}

  ]

\renewcommand*\rmdefault{bch}\normalfont\upshape
\rmfamily
\section*{}
\vspace{-1cm}


\footnotetext{\textit{$^{a}$~Delft University of Technology, Faculty of Mechanical Engineering, 2628 CD Delft, The Netherlands}}

\footnotetext{\textit{$^{\ast}$}~ Corresponding author. Email: p.fanzio@tudelft.nl}
\footnotetext{$^\dag$~ These authors contributed equally to this work}



\section*{Introduction}\label{intro}

Vascular fluidic phantom devices have become indispensable platforms for investigating fluid dynamics at scales relevant to physiological flows, enabling controlled studies of velocity fields, shear stress distributions, and the geometrical effects associated with vascular pathologies, such as aneurysms and stenoses \cite{Miao2025,Yalman2025}. Aneurysms, localized dilation of blood vessels, and stenoses, pathological narrowing, are of particular clinical interest because they dramatically alter local hemodynamics, influencing wall shear stress (WSS), flow separation, and the risk of vascular wall rupture or clot formation. Understanding these flow alterations is crucial to predict disease progression and to design therapeutic interventions. \\
One of the imaging techniques that enables the measurements of high-resolution velocity fields and transient flow behaviors that are otherwise challenging to measure is micro particle image velocimetry (micro-PIV)  \cite{ref:PIVGeneral}. Micro-PIV is a well‑established, non‑intrusive optical measurement technique that captures instantaneous velocity vector fields by tracking tracer particles seeded in the flow. The method is capable of resolving micron‑scale velocity distributions by illuminating tracer particles and computing displacements via cross‑correlation between image pairs. Successful micro-PIV experiments in microchannels have provided detailed insights into flow behavior under laminar conditions, enabling validation of analytical solutions such as Hagen–Poiseuille profiles as well as computational models \cite{etminan}.\\
Conventional fabrication techniques for vascular fluidic phantom, such as casting\cite{nilsson}, soft lithography, injection molding and hot embossing, lack sufficient geometric complexity, scalability, and reproducibility \cite{liu}. These constraints limit their ability to mimic anatomically realistic or even patient-specific vascular structures. 
Recent developments in 3D printing have overcome many of these challenges, enabling the fabrication of complex, 3-dimensional, patient-specific vascular geometries from medical imaging data \cite{ref:3DprintingVessels}. Techniques such as digital light processing (DLP) and stereolithography (SLA) enable the direct fabrication of devices with complex internal channels and  3D structures without the need for multi‑part assembly. Several articles review the use of 3D printing for the manufacturing of vascular fluidic phantom and microfluidic devices, highlighting its potential to overcome the limitations of traditional approaches and rapidly iterate on design models \cite{xu,gonzales}.\\
However, there are some limitations when using 3D printed vascular fluidic phantom. The minimum channel dimension is limited to the printer resolution and the ability to produce open, clean and defect-free channels. 
Moreover, achieving optical transparency and low surface roughness suitable for high‑quality optical flow measurements, like micro-PIV, remains challenging. Most commercial resins exhibit limited transparency and refractive index (RI) mismatches with typical working fluids, leading to significant light refraction and scattering at interfaces. Such effects degrade image quality, which depend on clear optical access to resolve particle location and reliably calculate the flow velocity. To address these challenges, researchers have pursued post‑processing techniques such as polishing and careful selection of materials and fluids for refractive index matching, to improve imaging performance in printed vascular fluidic phantoms \cite{a,b,c} .\\
In this study, we have demonstrated that 3D printing can be used to fabricate transparent vascular fluidic phantom devices with aneurysm and stenosis geometries down to feature sizes of $500 \mu m$. We further have shown that these devices are suitable for particle imaging velocimetry, upon appropriate post‑processing after 3D printing. We optimized a blood‑analogue working fluid by matching the refractive index with the printed material to minimize optical distortions and to enable reliable micro-PIV measurements. Starting with straight channel geometries, we performed micro-PIV experiments to compare measured velocity profiles with analytical solutions for laminar flow, validating both the manufacturing and measurement approaches. Building on this foundation, we investigated flow patterns in pathological geometries, highlighting the potential of this platform for experimental studies of vascular hemodynamics.

\section*{Materials and Methods}\label{MM}

\begin{figure*}
 \centering
 \includegraphics[height=10cm]{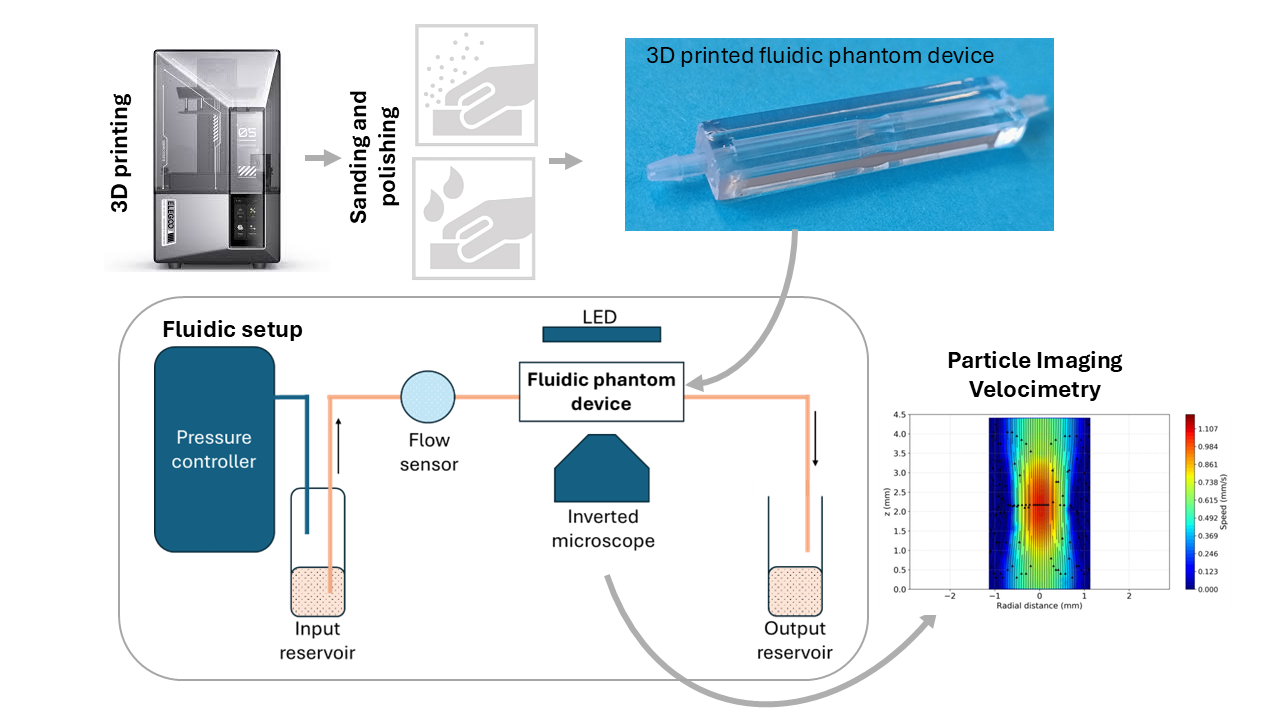}
 \caption{Schematic illustration of the manufacturing process, including 3D printing, post-processing (sanding and polishing), the fluidic setup used for flow measurement and resulting micro-PIV imaging.}
 \label{fig: PIV Experimental Setup}
\end{figure*}

\subsection*{Materials}\label{sec: Mat}
\textbf{The resin} used for the 3D printing of the microfluidic channels is Anycubic High Clear resin. 

\textbf{The working fluid} was developed to match the refractive index of the resin and it consisted of a mixture of water (29.01 wt\%), glycerol (12.91 wt\%), sodium iodide (58.02 wt\%), and sodium thiosulfate (0.058 wt\%) \cite{Jedrzejczak2023}. The dynamic viscosity of the working fluid was 7.15 mPa$\cdot$s. 
The working fluid exhibited Newtonian behaviour within the investigated flow rate range. Similarly, blood can be considered Newtonian in vessels of this size, with shear-thinning effects becoming relevant only in diameters below approximately 300~µm~\cite{Robin1931}. 
Density was determined by magnetically stirring the fluid for 5 minutes, followed by weighing three separate 1\,ml samples using a precision scale. The density was determined from the average of three measurements of a 1 mL sample, resulting in a density of 1.790 g/mL. 
For PIV measurements, the working fluid was seeded with solid silica particles (Bangs Laboratories) of mean diameter 2.47 $\mu$m. The seeding concentration was 5.5 mg in 15 ml of working fluid, corresponding to a weight-volume fraction of 0.04\% w/v.

\subsection*{Design of the fluidic phantoms devices}\label{sec: Design of Device}


\textbf{Straight fluidic} channels with circular cross section were designed having intended diameters of 0.5~mm, 1.0~mm and 2.0~mm. The 0.5~mm and 1.0~mm channels had a total length of 24 mm, whereas the 2.0~mm channel had a length of 29 mm. 

\textbf{Pathological models} were developed to represent a) aneurysmal and b) stenotic channels, each scaled by a factor of 1.25 relative to the nominal channel diameter of 2 mm, which in the stenotic geometry can be classified as a mild stenotic channel \cite{ClevelandClinic2022}. The widest section of the aneurysmal model is 2.5 mm, while the narrowest section of the stenotic model is 1.5 mm. The total length of these diseased models was 39 mm. 
\\
Each channel has a 1.0 mm diameter inlet and outlet nozzle. The corresponding nozzle designs are shown in Appendix \ref{D}.

\subsection*{3D printing of fluidic phantom devices}\label{sec: priting}

SOLIDWORKS was used to design the structures and create the stl files. Chitubox was used afterwards as the slicing software. The microfluidic devices were fabricated using an Elegoo Mars 5 Ultra masked stereolithography (MSLA) 3D printer. Different \textbf{printing settings} were evaluated to investigate the influence of layer height and printing orientation (appendix \ref{E}). The printing parameters are summarized in \autoref{tab:printing-settings}. 

\begin{table}[h]
\small
  \caption{\ Printing parameters used for Anycubic High Clear resin samples}
  \label{tab:printing-settings}
  \begin{tabular*}{0.48\textwidth}{@{\extracolsep{\fill}}lll}
    \hline
    \textbf{Layer Height} & \textbf{Exposure time} & \textbf{Bottom Exposure time} \\
    \hline
     50\,$\mu$m & 3.4\,s & 50\,s \\
    20\,$\mu$m & 2.3\,s & 40\,s\\
    \hline
  \end{tabular*}
\end{table}

In this paper, samples were printed at an angle (A) correspond to the pixel-to-layer ratio angle, $\theta_{pxl}$, determined to minimize the stair-stepping effect while printing under an angle \cite{deHaan2025}. This can be calculated using the equation $\theta_{pxl} = \arctan\left(\frac{\text{Layer Height}}{\text{X/Y Resolution}}\right)$, where the layer height is the printing layer height, and the X/Y resolution, the pixel resolution of the printer. This resulted in $\theta_{pxl}=$ 70.2$^\circ$ for the 50$\mu$m layer height and  $\theta_{pxl}=$ 48$^\circ$ for the 20 $\mu$m. Also, vertical (V) samples were printed at a 90$^\circ$ angle relative to the build plate. 

All samples underwent standardized \textbf{post-processing} steps to remove uncured resin. The procedure consisted of two baths in isopropyl alcohol (IPA), with the second bath sonicated (Ultrasonic Cleaner Tank, Model NA-4672, Shesto Ltd., Watford, UK), followed by UV curing for 20 minutes without heating using 405 nm UV light (Form Cure, Formlabs Inc., Somerville, MA, USA). 
To enhance optical transparency, additional post-processing steps were developed. Samples were wet-sanded for 30\,s per side using progressively finer grit sizes (P600 to P5000). Residual moisture and particles were removed with a microfiber cloth, after which the devices were coated with a thin layer of clear acrylic-based varnish (GAMMA Spuitlak Hoogglans Transparant 400~ l, product no.455016). The coated samples were left to dry for 18\,h in air. 

\subsection*{Morphological Characterization of Microfluidic Devices}\label{Morphology Measurements}

The evaluation of the \textbf{dimentional accuracy of the 3D printed microfluidic devices} was performed using the Keyence VHX-6000 Digital Microscope with 20×, 30x or 50x magnification and full ring illumination.  
By taking several images of the top and side view of the channel at different locations the error between the intended diameter and actual diameter was defined. This was calculated using the  equation $\text{D}_\text{{Error}} \% = \frac{1}{N} \sum_{i=1}^{N}(\frac{\lvert \text{D}_\text{{measured,i}}-\text{D}_\text{{designed}} \rvert}{\text{D}_\text{{designed}}}) \cdot 100\%$, with $\text{D}_\text{{Error}}$ being the relative error in percentage, $\text{D}_\text{{measured,i}}$ the $i^{th}$ measured diameter of either the top or side view, $\text{D}_\text{{designed}}$ the designed diameter of the channel, and N the number of measurements. 

In order to evaluate the \textbf{surface roughness} inside the channel at both printing orientations was estimated using a Nikon Eclipse TE2000-S inverted microscope with a Nikon Chromatic aberration Free Infinity (CFI) Plan Fluor 10x objective.  The 0.5 mm vertically printed channel and 1 mm angled printed channel were used for this analysis. Five peak-to-valley measurements were done on both samples, from which the surface roughness could be estimated for both printing orientations using the equation $R_Z = \frac{1}{N}\sum_{i=1}^{N}(H_\text{Peak-to-Valley,i})$, where R$_Z$ is the average peak-to-valley height, N is the number of measurements, and $H_\text{peak to valley}$ is the peak-to-valley height. 

\subsection*{Characterization of post processing and refractive index matching}\label{Sec: RI}
In order to measure the optical clarity,  a paper-printed 10~x~10~cm$^2$ background grid with a 1~mm square pattern has been used. \\
To assess the post-processing steps a $30 \times 30\,\text{mm}^2$ 3D printed plate was placed on top of the grid. To evaluate the refractive index matching, devices with a 2mm channel filled with air, water, the working fluid were placed on top of the grid. \\
In both cases, optical clarity was measured using ImageJ software. Grayscale profiles were extracted along an 8 mm line crossing four black and four white squares of the background grid. These profiles were compared with a control measurement taken from the grid without a sample. The deviation between the sample and the control grayscale values was computed, providing a quantitative measure of optical clarity.
The deviation in grayscale was calculated using the mean error. The mean error, $\text{G}_{\text{Mean Error}}$, represents the average grayscale deviation between a sample and the control and was calculated using equation: $\text{G}_{\text{Mean Error}} = \frac{1}{N} \sum_{i=1}^{N} \left( \frac{G_{\text{sample},i} - G_{\text{control},i}}{G_{\text{control},i}} \right) \cdot 100\% $, where N is the total number of pixels, $G_{\text{sample},i}$ the grayscale value of the sample at position i, and $G_{\text{control},i}$ the grayscale value of the control at position i.

\subsection*{Micro-PIV and flow rate measurements}\label{Sec: flow rate and Pdrop}

The experimental setup (as shown in the schematic in figure\ref{fig: PIV Experimental Setup}) consists of a pressurized input reservoir to drive the flow, the microfluidic device, a flow sensor, and an output reservoir open to atmospheric pressure. The channel is illuminated by an LED light and visualized using an inverted microscope. 
Flow control is achieved with a Fluigent FlowEZ flow controller, driven using a FLPG+ Pressure Supply Line (0–2.3\,bar range) and a flow sensor, Fluigent Flow Unit L+ (0–40\,ml/min range).
The flow rate sensor was factory-calibrated for water or IPA, a manual recalibration was required to ensure accurate measurements with the tracer-seeded working fluid. A detailed overview of the calibration process can be found in Appendix \ref{F}. 

Connections were made using fluorinated ethylene propylene (FEP) tubing (1/16” OD, 0.020” ID = 508 $\mu$m) and silicone tubing (2.1 mm OD, 0.5\,mm ID). The tight fit between the tubing and device ports ensured a leak-free connection, eliminating the need for glue or other sealing agents. 
Before each measurement, trapped air was removed by setting the inlet pressure to 2.0 bar and holding the device vertical and by switching the flow direction until a stable pressure drop was obtained at the set flow rate. After each measurement, the flow loop was flushed with 20 ml of IPA followed by 40 ml of distilled water. 

Micro-PIV measurements were performed using the Nikon Eclipse TE2000-S inverted microscope with a Nikon CFI Plan Fluor 10× objective (N.A. 0.3, W.D. 16 mm) and a Nikon CFI Plan Fluor 4× objective (N.A. 0.13, W.D. 17.2 mm). A continuous LED source illuminated the microfluidic device. Images were recorded with a Nova S12 Fastcam, providing a resolution of 1024 × 1024 px$^2$ at up to 12,800 frames per second (FPS), with a 20 $\mu$m pixel size and 12-bit depth. The field of view (FOV) ranged from 2048 × 2048 $\mu$m$^2$ for the 10× objective to 5120 × 5120 $\mu$m$^2$ for the 4× objective. 

For each measurement, 3000–10,000 consecutive image pairs were recorded. The frame rate was adjusted to ensure maximum particle displacements of approximately 10–15 pixels. 
Recordings were performed at five steady-state flow rates: 65, 150, 240, 535, and 950 $\mu$L/min. 

\subsection*{Data Analysis}\label{sec: data analysis}

\subsubsection*{Analysis of micro-PIV Data}\label{subsec: Data Analysis Micro-PIV} 

Post-processing of micro-PIV data was performed using DaVis (LaVision GmbH) and Python 3.6. The raw images were inverted using the invert buffer option in DaVis to produce light tracer particles against a dark background, improving contrast for correlation analysis. Since the entire device was illuminated with continuous LED lighting, a mask was applied to exclude regions outside the channel, ensuring that only the fluid domain was analyzed.  

The correlation cutoff, $\epsilon$, is chosen as 0.1, consistent with the value used in previous studies \cite{Wereley2005}. This results in a Depth of Correlation (DOC) of 64.65 $\mu$m for the 10x objective and 344.92 $\mu$m for the 4x objective. The DOC represents 12.9$\%$ of the channel height for the 500 $\mu$m channel, 6.4$\%$ of the 1000$\mu$m channel and only 3.2$\%$ for the 2000 $\mu$m channel at 10x. For the 4x objective, the DOC represents 17.3$\%$ for the 2000 $\mu$m channel and 34.5$\%$ for a channel of 1000 $\mu$m in diameter. 

Micro-PIV analysis was conducted using a multi-pass cross-correlation algorithm. Interrogation windows of 64 × 64 pixels with 50$\%$ overlap were applied initially, followed by two refinements to 32 × 32 pixels with 50$\%$ overlap. Vector validation was performed to ensure data quality. Vectors were deleted when the correlation peak ratio ($Q$) was below 1.2. Universal outlier detection was applied with a residual threshold of 2.0 and a 5 × 5 filter region. Rejected vectors were reinserted when the residual was below 3.0 with a minimum requirement of four neighboring vectors. Groups with fewer than five vectors were removed, and empty spaces were interpolated while preserving validated vectors.  

The physical velocity components,  $u$ and $v$, were obtained using the equations:
\begin{equation}\label{eq:VelocityFromDisplacement}
u = \frac{\Delta x \cdot px_{size}}{\Delta t \cdot M}, \qquad v = \frac{\Delta y \cdot px_{size}}{\Delta t \cdot M}
\end{equation}

where $M$ is the magnification, $px_{size}$ is the pixel size of the camera sensor in mm, $\Delta t$ the time between the consecutive frames in seconds and $\Delta x$ is the displacement. 

Velocity fields were obtained using ensemble correlation, where multiple image pairs are correlated together to provide converged flow fields. From these flow fields, the radial velocity profiles were then extracted. For straight channels, profiles were averaged along the channel length, while for pathological channels, profiles were computed within selected regions of interest (ROI), such as the most constricted or dilated sections in the aneurysmal and stenotic design. The radial velocity profile was calculated using the equation:
\begin{equation}\label{eq:RadialProfile}
u_r(r) = \frac{1}{L} \int_0^L u(r)\, dz
\end{equation}

where $u_r(r)$ is the radial velocity at distance $r$ from the channel center, averaged along the axial length $L$ or ROI length. The standard deviation of the recorded radial velocity was also calculated and used to estimate the error.

From the velocity gradient at the channel wall, the wall shear stress (WSS) were calculated using the equation:
\begin{equation}\label{eq:WSS}
\tau_w = \mu \left. \frac{\partial u}{\partial r} \right|_{r=R}
\end{equation}

where $\tau_w$ is the WSS, $\mu$ is the dynamic viscosity of the working fluid, and $\partial u / \partial r$ is the radial velocity gradient evaluated at the channel wall ($r = R$).  

Lastly, contour maps of velocity magnitude with superimposed streamlines to visualize flow organization were generated from the PIV data using Python 3.6.  

\subsection*{Analytical Analysis}\label{sec: analytical and computational approach}

The analytical model was used only to the straight channel designs. Hagen–Poiseuille was applied under the assumptions of steady, laminar, and fully developed flow of an incompressible Newtonian fluid in a straight circular channel of constant radius. A no-slip boundary condition was imposed at the wall, and fluid properties were assumed constant at uniform temperature. 
Considering $L$ as channel length, $R$ as channel radius, $Q$ as volumetric flow rate and $\mu$ as the dynamic viscosity of the fluid, the pressure drop, $\Delta P$, across the channel is calculated using the equation:
\begin{equation}\label{eq:pressuredrop}
\Delta P = \frac{8 \mu L Q}{\pi R^{4}}
\end{equation}
The radial velocity profile is estimated with the equation: 
\begin{equation}\label{eq:prof}
u(r) = \frac{2Q}{\pi R^2} \left(1 - \frac{r^2}{R^{2}}\right)
\end{equation}
where $u(r)$ is the velocity at distance $r$ from the channel centerline.
The wall shear stresses (WSS), $\tau_{w}$ , were calculated using the equation: 
\begin{equation}\label{eq:WSS}
\tau_{w} = \frac{4 \mu Q}{\pi R^{3}}   
\end{equation}

This analytical model was compared with the experimentally measured pressure drop, radial velocity profiles and WSS obtained from micro-PIV analysis to assess its validity.

\begin{figure*}[bth]
 \centering
 \includegraphics[height=10cm]{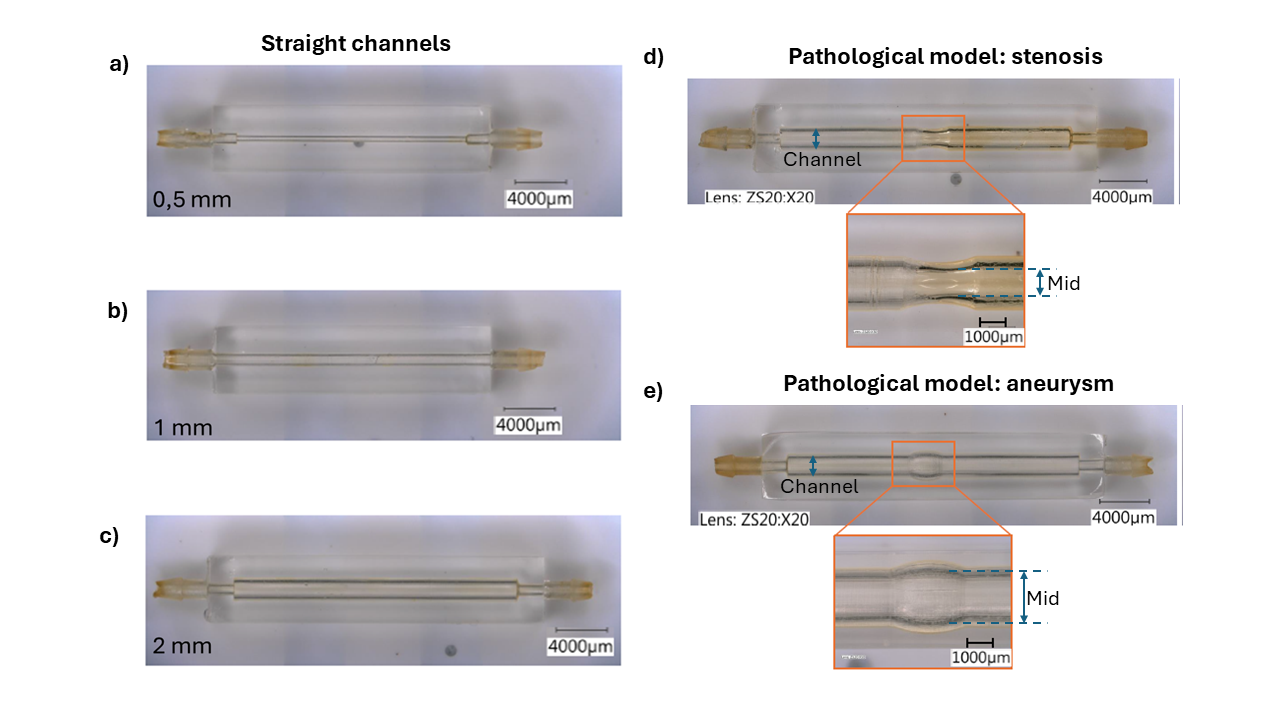}
 \caption{3D-printed straight channels with diameter of (a) 0.5 mm, (b) 1.0 mm, and (c) 2.0 mm channel. 3D printed diseased channels replicating (d) a stenotic channel, and (e) an aneurysmal channel.}
 \label{fig: Morphology Straight}
\end{figure*}

\section*{Results and Discussion}

\subsection*{3D printing of fluidic phantom devices}

\begin{figure}
\centering
\includegraphics[width=1\linewidth]{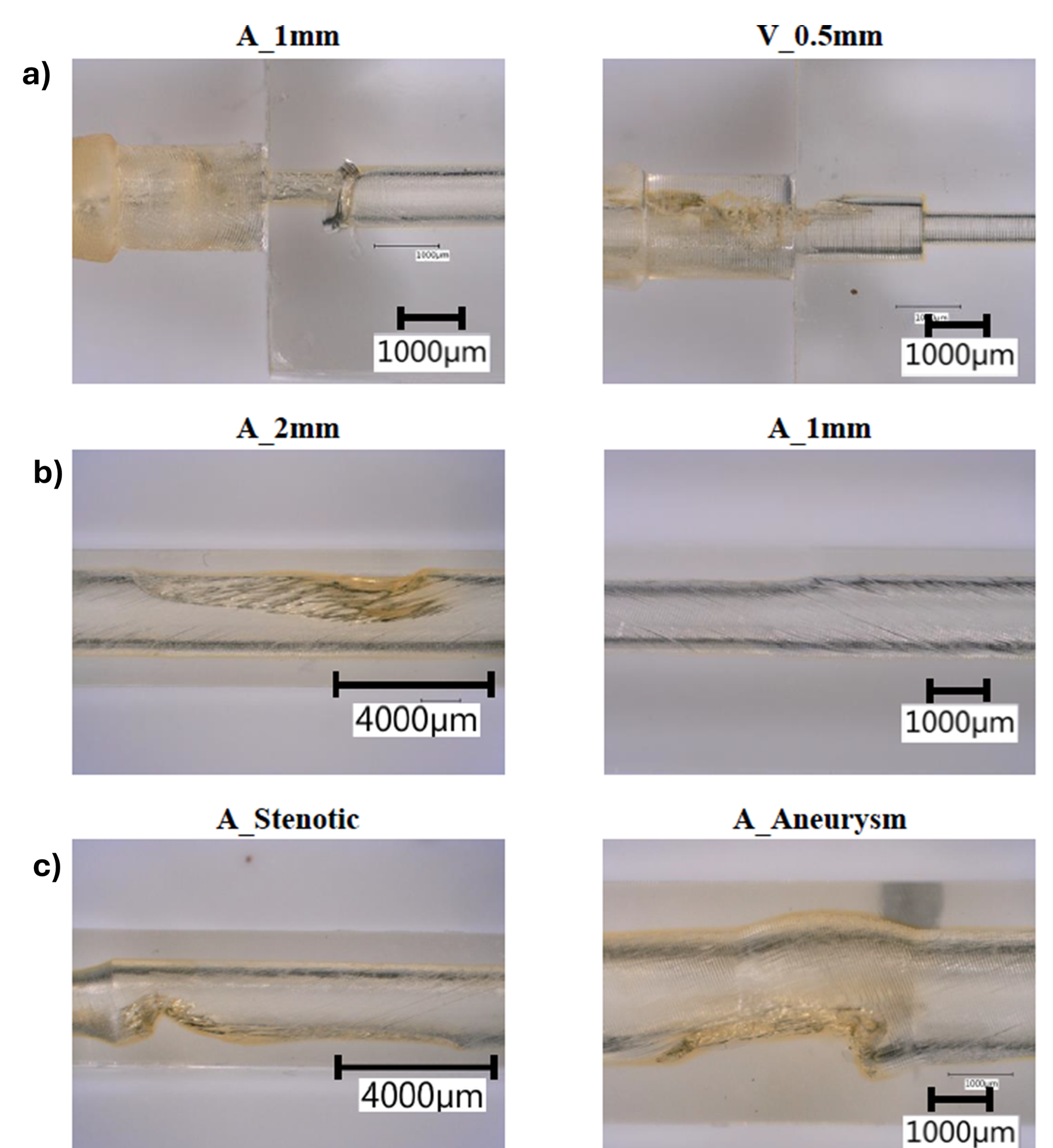}
\caption{Examples of 3D printing failure modes. Defects causing diameter reduction at the inlet nozzle location in a) a 1mm diameter channel printed at an angle and b) a 0.5mm diameter channel printed vertically. Localized channel narrowing in a 2mm diameter c) and in a 1mm diameter channels d) both printed at an angle. Artifacts in angled stenotic e) and aneurysmal f) channels. The 'V' relates to vertically printed and 'A' to printed at an angle.}
\label{fig: Failing modes 3D printing} 
\end{figure}

Figure \ref{fig: Morphology Straight} shows 3D-printed fluidic devices featuring straight channels of 0.5mm (a), 1mm (b) and 2mm (c) and 2 diseased models replicating a stenosis (d) and an aneurysm (e).
The evaluation of the accuracy of the 3D printing process is presented in table \ref{tab:dimensional_straight}. All target diameters for the straight channels are reproduced with relative errors ranging from $0.13\%$ to $0.57$\% . This confirms that the 3D printing and post-processing preserve the intended diameters across the straight channels. For the diseased channels results are similar, with relative errors ranging from $0.37\%$ to $1$\%. However, the manufacturing of the middle part of the aneurysm shows a relative error of 17\%.
These location-specific artifacts align with regions where the cumulative UV dose is higher \cite{WANG2024104350}. This effect is particularly pronounced near sharp geometric transitions or in angled orientations because of unintended recurrent exposure of resin in the case of angled printing specifically. In addition, there is a possibility of residual polymerization which is exacerbated at relatively narrow sections of the structure. \cite{WANG2024104350} 
Figure \ref{fig: Failing modes 3D printing} summarizes the printing limitations and failure modes observed across the devices. Figures \ref{fig: Failing modes 3D printing} (a) and (b) illustrate common inlet and outlet nozzle defects, characterized by an apparent reduction in nozzle diameter. In both images, the nozzle failed due to resin accumulation near the channel end, which subsequently cured, consistent with the overcuring mechanism described above. This effect is amplified when printing at an angled print orientation (figures \ref{fig: Failing modes 3D printing} (a)). 
In the channel sections, printing errors appeared as either abrupt narrowing (figure \ref{fig: Failing modes 3D printing} (c) ) or subtle diameter variations along the length (figure \ref{fig: Failing modes 3D printing} (d)). The abrupt changes are consistent with local overcuring if uncured resin accumulates in presence of small local defects, whereas the more gradual variations can also arise from slight misalignment of the device with respect to the build plate during the print process.

\begin{table}[H]
\centering
\caption{Summary of mean of dimensional accuracy for the 3D- straight channels.} 
\label{tab:dimensional_straight}
\begin{tabular}{llcc}

\hline
\textbf{Printing } & \textbf{Channel} & \textbf{Designed diameter} & \textbf{D{$\_Error$}}\\
\textbf{Orientation} & \textbf{Type} & \textbf{(mm)} & \textbf{$\%$}\\
\hline
Angled (A) & Straight & 0.5 & 0.57 \\
  & Straight & 1.0 & 0.52\\
  & Straight & 2.0 & 0.36  \\
  & Stenotic & mid & 0.37\\
 & Stenotic & channel & 0.78\\
 & Aneurysm & mid & 17\\
 & Aneurysm & channel & 0.9\\
\hline
Vertical (V) & Straight & 0.5 & 0.27  \\
  & Straight & 1.0 & 0.57 \\
  & Straight & 2.0 & 0.13  \\
   & Stenotic & mid & 1\\
 & Stenotic & channel &0.26\\
 & Aneurysm & mid & 0.76\\
 & Aneurysm & channel & 0.46\\
\hline
\end{tabular}

\end{table}

For the angled pathological models (Figures \ref{fig: Failing modes 3D printing} (e) and (f)), changes in the diameter of base channel led to overcuring within the pathological segment for the aneurysmal model and directly downstream of the constriction for the stenotic model, similar to the nozzle effect observed for the 0.5~mm channel (Figures \ref{fig: Failing modes 3D printing} (b)). \\
In conclusion, overcuring is therefore one of the main challenges when printing channels under an angle or with varying diameters. It can be mitigated by optimizing exposure parameters to reduce cumulative UV dose \cite{WANG2024104350} or by compensating for expected overcuring in the design stage \cite{deHaan2025}. Comparing the results obtained printing the devices vertically or at an angle, we can confirm that both the orientation leads to acceptable print quality. However, in our experience, printing the device angled gives more chances of having failed or defect prints.\\
In this work, all the devices showing the local defects described previously have not been used for subsequent micro-PIV measurements.

\subsection*{Surface Roughness}

\begin{figure}[H] 
\centering
\includegraphics[height=5cm]{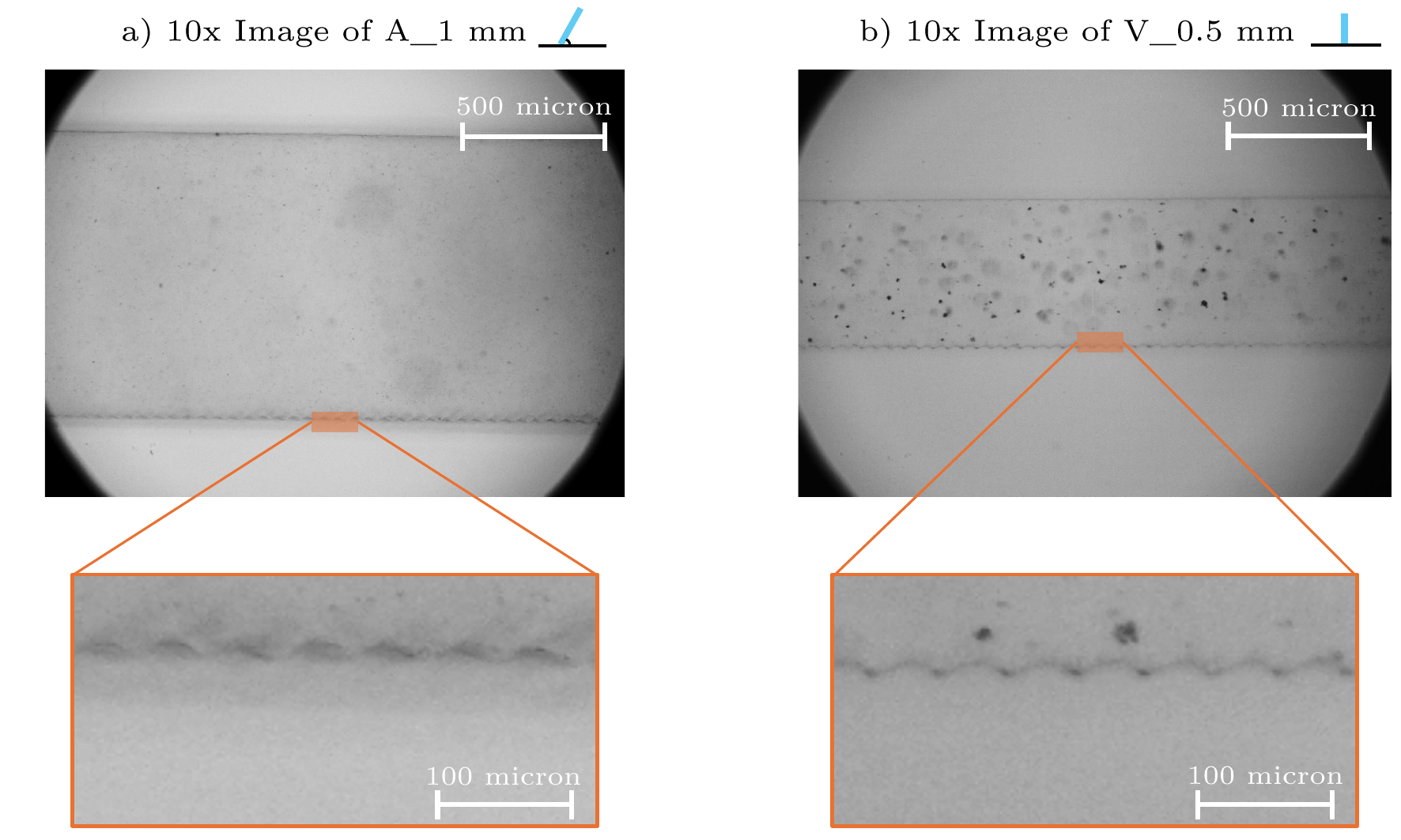}
\caption{Images showing the surface texture on the inner walls of the channel of a) a vertically and b) an angled printed 1mm diameter channel. The roughness along the wall is clearly visible in both samples.}
\label{fig: Surface Roughness PIV}
\end{figure}

Analysis of the surface roughness for both vertically and angled printed channels was performed using images taken with the high-speed camera through the inverted microscope at 10x magnification. Figure \ref{fig: Surface Roughness PIV}  shows representative images of the 1~mm angled channel and the 0.5~mm vertically printed channel, including zoomed-in insets of the wall regions where surface features are most visible.
Based on these images, the peak-to-valley height ($R_Z$) was estimated as 13.7~$\pm$~5.7~$\mu$m for the vertically printed channel and 12.6~$\pm$~2.3~$\mu$m for the angled printed channel. The measurements were taken at five locations for each sample, providing a rough estimate of the surface roughness across the channel. The results indicate a minor difference between the two printing orientations, although the limited number of measurements makes it difficult to draw definitive conclusions.
However, the reported $R_Z$ values are significantly smaller than the printed channel diameter and therefore we can assume that roughness does not affect the flow measurements. 

\subsection*{Optimization of device transparency and refractive index matching}
This section presents the optical characterization of the microfluidic devices and the working fluid. First, the results of the developed post-processing procedure are discussed, followed by the evaluation of refractive index (RI) matching between the working fluid and the resin.\\

\begin{figure}[H]
\centering
\includegraphics[width=0.9\linewidth]{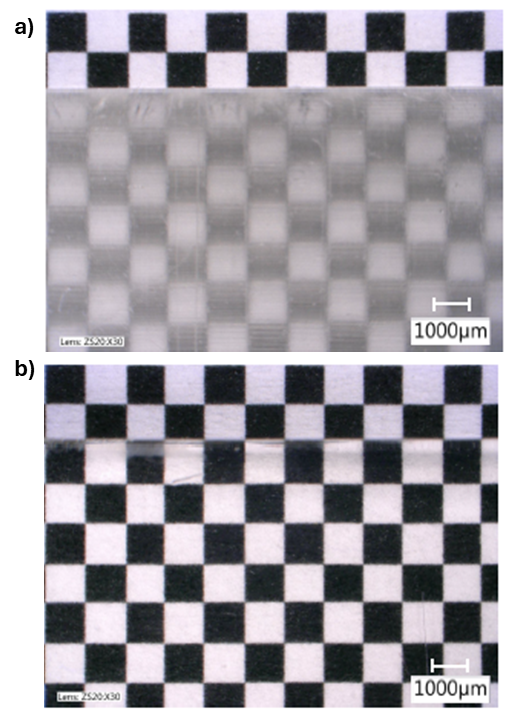}
\caption{Effect of additional post-processing steps on 30x30 mm$^2$ 3D printed plates: (a) untreated sample after curing, and (b) sample after wet-sanding and varnish coating, showing improved transparency and reduced distortion.}
\label{fig: Post-Processing Results}
\end{figure}

Figure \ref{fig: Post-Processing Results} compares an untreated 30×30 mm$^2$ sample printed at an angle (a) with a wet-sanded and varnish-coated sample (b). The additional post-processing steps clearly enhance optical transparency, substantially reducing visual distortion across the sample. Both orientations yielded similar transparency after complete treatment. 
To further assess printing orientation effects, three orientations, vertical, horizontal and angled were analyzed quantitatively. Table \ref{PP} shows that the mean grayscale errors were small and comparable in all orientations, with a maximum deviation of 5.7\% for the vertically printed sample. These results indicate that printing orientation has only a minor influence on optical transparency once post-processed.\\

\begin{figure}
\centering
\includegraphics[width=0.8\linewidth]{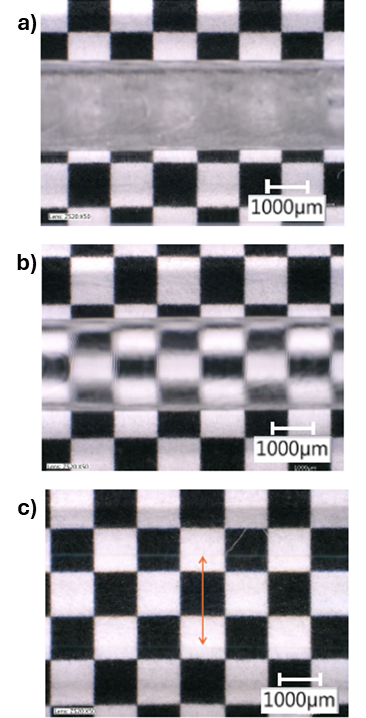}
\caption{Demonstration of the effect of refractive index in a 2 mm channel with different media: (a) air , (b) distilled water, and (c) the developed working fluid.}
\label{fig: WF}
\end{figure}

In addition to post-processing, accurate micro-PIV measurements require optically clear models and index-matched working fluids, making the choice of working fluid a critical step \cite{Jedrzejczak2023}. In \textit{in vitro} vascular research, blood-mimicking fluids (BMF) are commonly used to replicate the density and dynamic viscosity of blood \cite{Bordones2018}. For micro-PIV measurements, however, the RI of the working fluid must match the RI of the model's material to ensure accurate results \cite{ref:PIVGeneral}. If the RI values do not match, the light can refract and reflect at the interface between the fluid and the model. This can lead to image distortion, blurring, and displacement errors, which cause measurement errors.
PDMS devices (RI $\approx$ 1.41) are often paired with water–glycerol mixtures \cite{Doutel2015, Yadzi2019, Hong2017}. Resin-printed devices usually have higher RI ($\sim$1.50–1.51), motivating the addition of RI-raising additives (e.g. sodium iodide, Ammonium thiocyanate) \cite{Ho2020, Gallagher2018, Jedrzejczak2023}. 
In this work, sodium iodate and sodium thiosulfate have been added to water. These additives can be hazardous and alter density/viscosity, making it challenging to simultaneously match viscosity, density, and RI.  For this reason, dynamic similarity within the physiological range was achieved using Reynolds numbers between 2 and 6. Although no definitive Reynolds number values exist for physiological conditions, the selected range is consistent with physiological behavior. 
The RI matching  is demonstrated in figure \ref{fig: WF}: as the RI of the medium increases, from air ($\approx$ 1.00) (a) to water ($\approx$ 1.33) (b) to the developed working fluid ($\approx$ 1.50) (c), the image distortion decreases. With the RI-matched working fluid, only the outer channel wall remains visible and the background grid appears undistorted, confirming near-perfect optical matching.  

Overall, these results demonstrate that the Anycubic High Clear resin exhibits an RI well matched to the developed working fluid, producing clear, distortion-free visualization suitable for accurate micro-PIV measurements. 

\begin{table}[H]
\centering

\caption{Normalized mean grayscale error for samples printed at different orientations after wet-sanding and coating. N denotes the number of pixels used for the analysis.}
\label{PP}
\begin{tabular}{lcc}
\textbf{Sample orientation}& \textbf{G}$_{\textbf{Mean Error}}$ [$\%$] & \textbf{N} \\
\hline
Vertical& 5.67 & 1005 \\
Horizontal& 4.96 & 1114 \\
Angled& 4.58& 1114 \\
\hline
\end{tabular}
\end{table}

\subsection*{Flow measurement in straight channels}

In this section, the micro-PIV results of the straight channels are discussed, including the radial velocity profile, a contour plot with streamlines, and the wall shear stress. 

\begin{figure*}
 \centering
 \includegraphics[height=12cm]{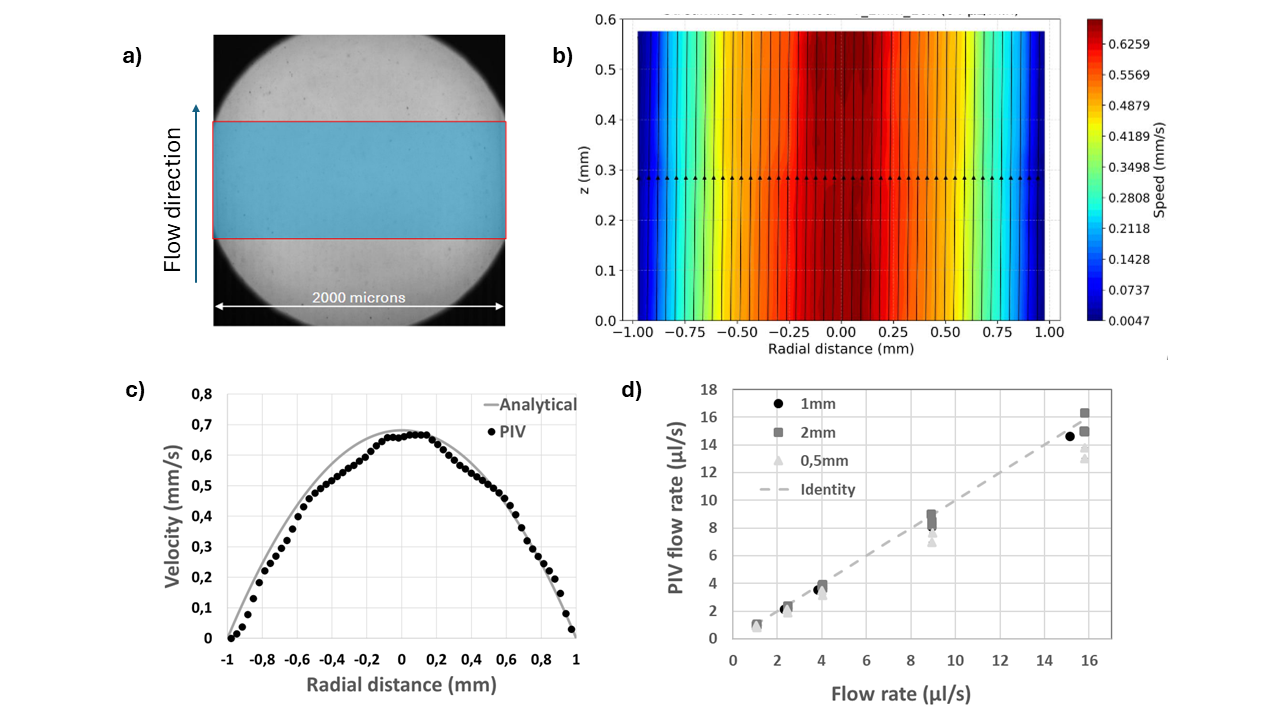}
 \caption{micro-PIV measurement on a 2mm diameter straight channel at a flow rate of $1 \mu l/s$: a) raw data, b) contour plot and c) velocity profile. d) Comparison between the flow rate measured using PIV and the flow rate sensor for 3 channel diameters.}
 \label{fig: Micro-PIV a_2_2nd_1uls}
\end{figure*}

The micro-PIV measurement in a 2 mm diameter channel at 1\(\mu\)L/s is presented in figure \ref{fig: Micro-PIV a_2_2nd_1uls}: the raw micro-PIV image with the investigated region in the blue square (a) and the contour plot (b). Figure  \ref{fig: Micro-PIV a_2_2nd_1uls} c) shows that the radial velocity profile obtained from the micro-PIV (dots) matches well with the analytical model (line).\\
The plot in  \ref{fig: Micro-PIV a_2_2nd_1uls} d) summarizes all the flow rates measured using the PIV velocity profiles for the 3 channel diameters (0.5mm, 1mm and 2mm) versus the flow rate measured with the flow sensor. There is a good agreement between the values, demonstrating that it is possible to perform PIV measurements using the 3D printed devices. 
Across all the measurements preformed at different flow rate and channel diameter, we notice that the mean relative error in the velocity estimation from the PIV data compared with the analytical model ranges between 5\% to 17\%. 
There are several reasons that can explain such variations:
\begin{enumerate}
    \item  particles adhering to the wall, which can cause a locally reduced velocity magnitude, thereby increasing the mean error \cite{daaboul2023testing}.
    \item contamination on the camera sensor itself or on the surface of the microfluidic device, leading to a difference in contrast or obscuring the visualization. This can consequently result in a smaller velocity vector, which in turn leads to a higher mean error.
    \item  calibration imperfections of the camera sensor, which introduced periodic bands of darker and lighter intensity \cite{Michael}.
    \item   the depth of correlation averaging results in a smaller velocity magnitude~\cite{Kloosterman2011}, thereby underestimating the velocity magnitude.
    \item The mean relative error increases towards the walls of the channel, which is most probably caused by the difficulty in tracking particles close to the boundary. One of the reasons is that the wall curvature can cause measurement errors, as it can lead to light refraction and local distortion \cite{Yi2022}. 
    \item Some raw micro-PIV images exhibit subtle grayscale variations across the diameter, which can bias the cross-correlation and lead to periodic artifacts.
    \item  a slight optical distortion is also introduced due to minor refractive index differences between layers. The exact influence of this distortion on the resulting PIV data remains unclear but may contribute to the oscillating patterns observed in some measurements.
    \item If the pixel displacement exceeds approximately one-quarter of the interrogation window, aliasing of the correlation peak can occur \cite{Westerweel1990}. Even though displacement decreases near the walls, overlap between interrogation windows may cause periodic wrapping, where the displacement of tracer particles at the edge of the IW between two frames is misinterpreted, resulting in a false velocity vector. Zero padding, used in PIV processing, adds a border of zeros around the interrogation window to reduce edge effects during cross-correlation, preventing this pattern from spreading to the walls \cite{Westerweel1990}. 
\end{enumerate}

As the channel diameter increases, the spatial resolution improves, resulting in a greater number of velocity vectors across the channel cross-section. This leads to a reduction in the relative error, as each vector represents a smaller section of the total diameter, therefore minimizing the spatial averaging. This is observed when comparing the mean relative error between the different channels, with the greatest error for the 0.5~mm channel (14.6\%), followed by the 1.0~mm channel (7.81\%), and finally the 2.0~mm channel (6.3\%). 

The relative influence of surface roughness is greater in smaller channels because the ratio between the peak-to-valley height and the channel diameter is larger. However, all measurements were conducted under laminar conditions (maximum Reynolds number $\approx 10$). For fully developed laminar flow of a Newtonian fluid in a circular tube, the velocity profile and the increased friction factor are independent of wall roughness; therefore, roughness has a minor effect on the velocity profile at steady-state flow \cite{moody1944friction, White2006}.

The WSS was determined from the slope of the velocity profile near the wall and compared with analytical predictions in table \ref{tab:wss_summary_allchannels_mpa}. Under laminar flow and low Reynolds number conditions, the WSS should closely follow the smooth-wall approximation \cite{White2006, moody1944friction}. For the 0.5~mm, 1.0~mm, and 2.0~mm channels, the experimentally determined WSS values are notably lower than the analytical estimates, with a maximum relative error of $33.8\%$ for the vertical 0.5~mm channel and a minimum error of $12.3\% $ for the vertical 1~mm channel. The vertically printed channels generally show higher WSS than the angled ones for the 1~mm and 2~mm diameters, while the opposite is true for the 0.5~mm channels, making it difficult to identify a consistent orientation trend.

\begin{table*}
\small
  \caption{Summary of the average wall shear stress (WSS) values obtained from PIV data for all straight channels, compared with analytical predictions. }
  \label{tab:wss_summary_allchannels_mpa}
 \begin{tabular*}{\textwidth}{@{\extracolsep{\fill}}lllll}
    \hline
    \textbf{Channel diameter} & Orientation & Magnification & \textbf{Mean Measured WSS}  & \textbf{Analytical WSS} \\
    mm & & & $\times10^{-4}$ [Pa] & $\times10^{-4}$ [Pa]  \\
    \hline
    0.5 & Angled & 10x & 26.90 $\pm$ 1.86 & 37.60  \\
    0.5 & Vertical & 10x & 25.10 $\pm$ 0.92 & 37.60  \\
    \hline
    1 & Angled & 10x & 1.43 $\pm$ 0.03  & 1.76   \\
    1 & Vertical & 10x & 3.96 $\pm$ 0.07  & 4.62   \\
    \hline
    2 & Angled & 10x & 0.58 $\pm$ 0.03  & 0.59   \\
    2 & Angled & 4X & 0.52 $\pm$ 0.00  & 0.59   \\
    2 & Vertical & 10x & 0.76 $\pm$ 0.04  & 0.59   \\
    \hline
  \end{tabular*}
\end{table*}

\subsection*{Diseased Models}
\begin{figure*}
 \centering
 \includegraphics[height=10cm]{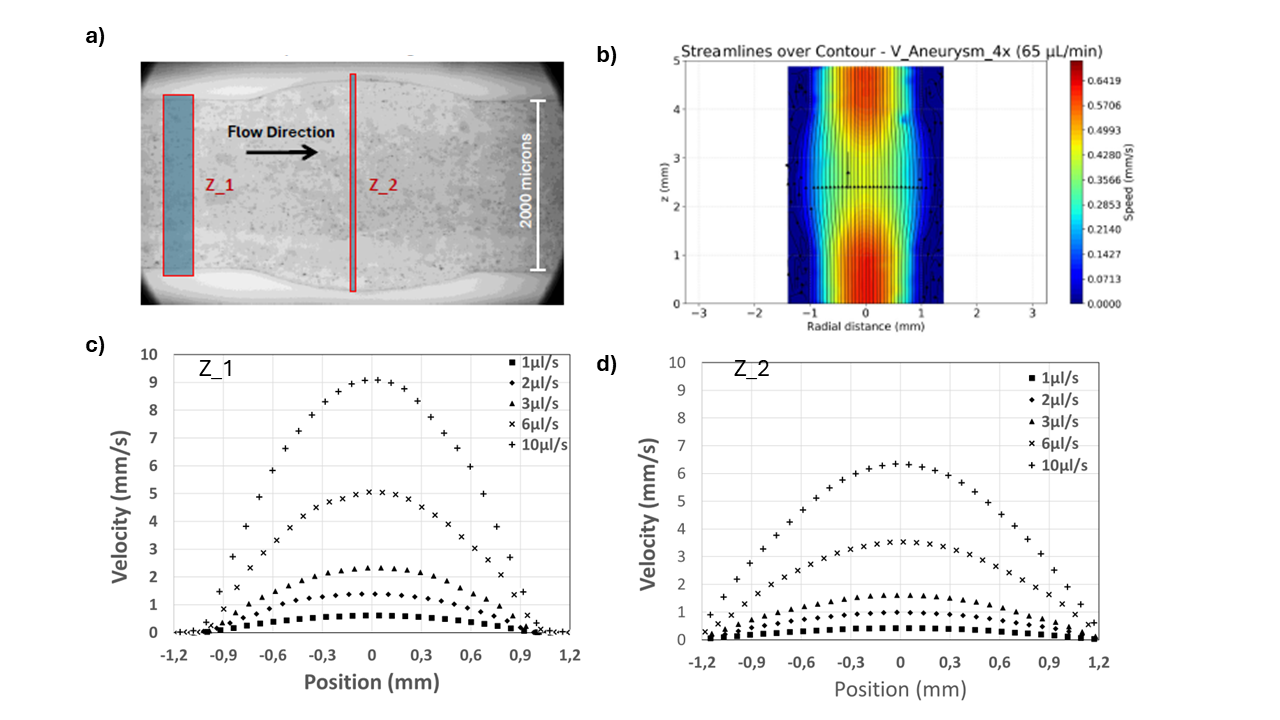}
 \caption{micro-PIV measurement of an aneurysmal model at a flow rate of $1 \mu l/s$: a) raw data, b) contour plot. c) Velocity profile at different flow rates in $Z_1$ region. d) Velocity profile at different flow rates in $Z_2$ region.}
 \label{fig: AN}
\end{figure*}

\begin{figure*}
 \centering
 \includegraphics[height=10cm]{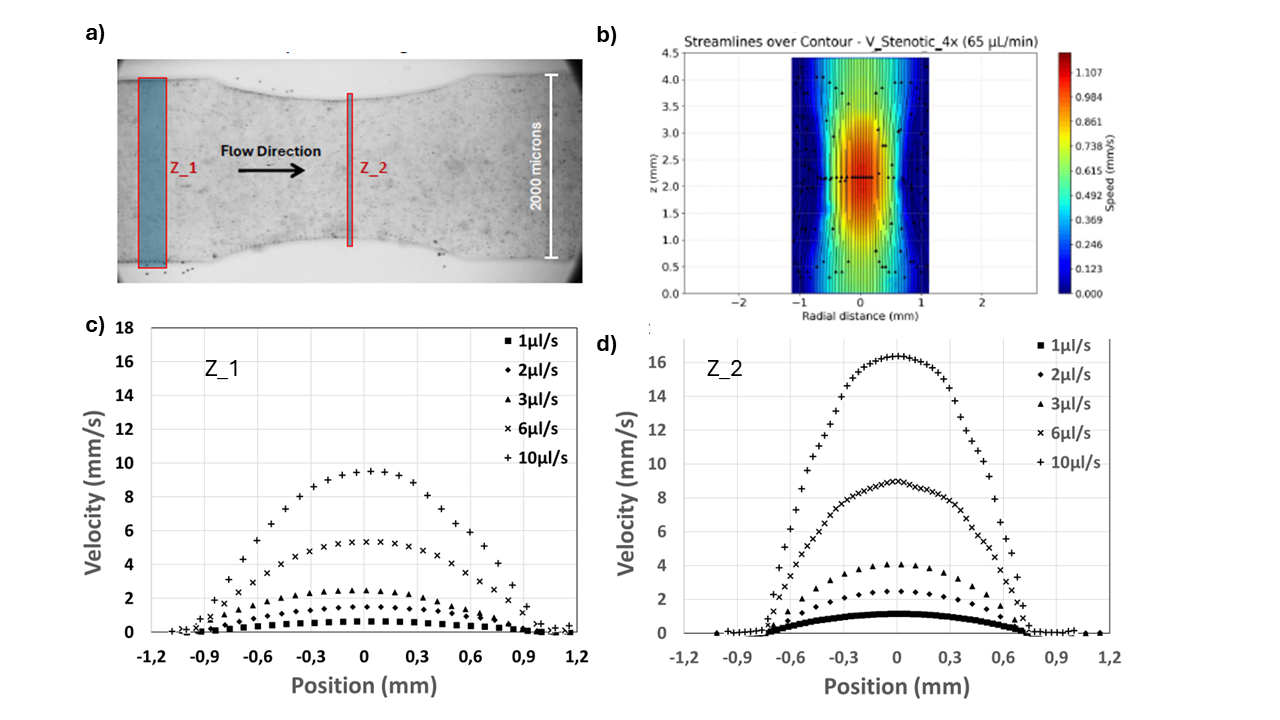}
 \caption{Particle imaging velocimetry measurement of a stenotic model at a flow rate of $1 \mu l/s$: a) raw data, b) contour plot. c) Velocity profile at different flow rates in $Z_1$ region. d) Velocity profile at different flow rates in $Z_2$ region}
 \label{fig: ST}
\end{figure*}

Figures \ref{fig: AN} (a) and \ref{fig: ST} (a) show the raw micro-PIV image of the aneurysmal model and the stenotic model. Figures \ref{fig: AN} (b) and \ref{fig: ST} (b) show the contour of  the aneurysmal model and the stenotic model. For both figures, the flow rate was set as for $1\mu l/s$, as rapresentative micro-PIV data. 

To obtain a more detailed view of the local flow dynamics, two regions were selected. Region Z\textsubscript{1} is located upstream of the diseased area, while region Z\textsubscript{2} is positioned within the narrowest section of the channel for the stenotic model and in the widest region for the aneurysmal model. \\
For the aneurysmal model, the velocity profiles in the Z\textsubscript{1} region are shown in figure  \ref{fig: AN} c) and the velocity profiles in the Z\textsubscript{2} region are shown in figure  \ref{fig: AN} d). As expected, for all the tested flow rates, the local velocity decreases as the channel widens and increases again downstream.  Due to the large field of view (FOV) required, both regions were analyzed using the 4× objective. 

For the stenotic model, the velocity profile in region Z\textsubscript{1} (figure \ref{fig: ST} (c)) is slightly asymmetric. In contrast, the profile in region Z\textsubscript{2} (figure \ref{fig: ST} (d)) is symmetric and, as expected, shows an increase in speed in the stenotic region for all the tested flow rates. 

\begin{figure}[H] 
\centering
\includegraphics[width=1\linewidth]{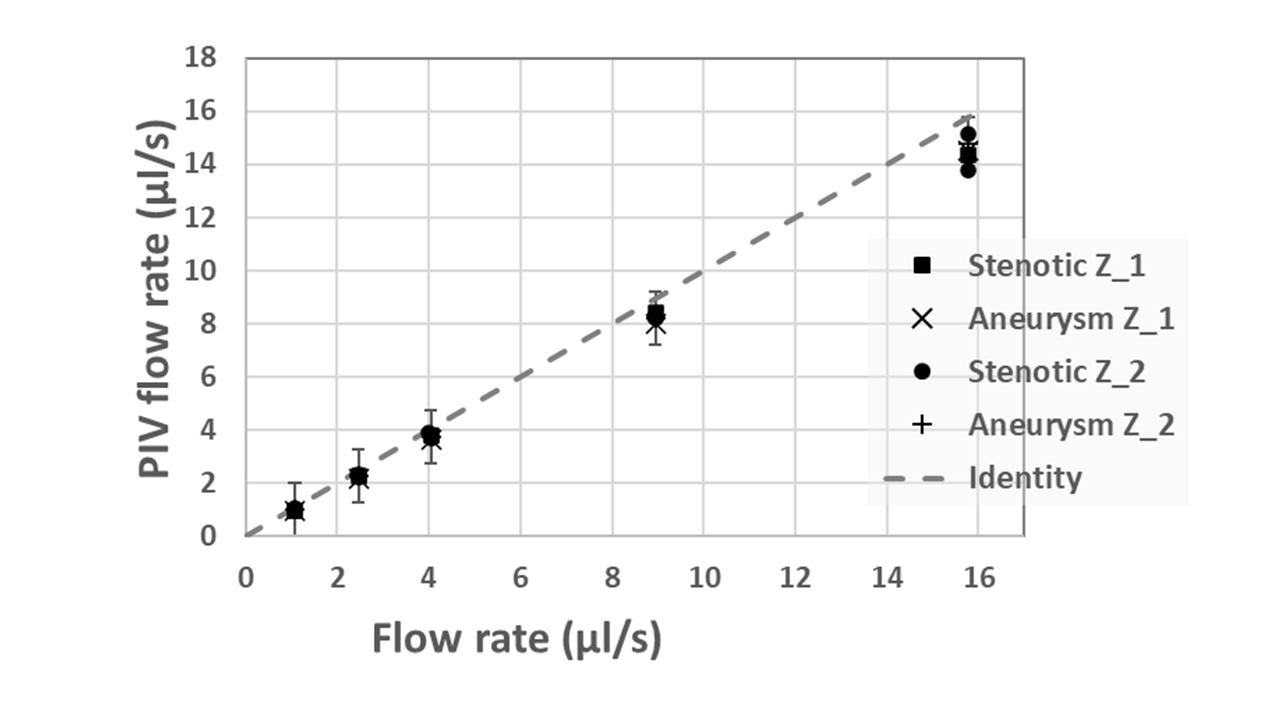}
\caption{Comparison between the flow rate measured using PIV and the flow rate sensor for the stenotic and the aneurysmal models}
\label{fig: SUM}
\end{figure}

Figure \ref{fig: SUM} a) shows that the average flow rate calculated from the PIV matches very well  the one measured by the flow rate sensor at low flow rates. \\

\section*{Conclusion}
This study demonstrates that, using an MSLA 3D printer, straight, aneurysmal and stenotic, channels with minimum diameters of 500~µm can be successfully fabricated. Vertically oriented printing proved most effective in preserving channel geometry accuracy and minimizing print errors.
Overcuring was identified as the most limiting factor in the fabrication of the fluidic channels. This effect can be reduced by optimizing or lowering the cumulative exposure dose and by adapting the channel design to compensate for expected overcuring. The printed nozzle design allowed for easy connection to the flow loop without leakage, although a reduction in inner diameter was observed, particularly for the 0.5~mm channel, which was again attributed to overcuring.

A post-processing method combining wet-sanding and varnish coating was developed,  which significantly improved the optical transparency of the printed devices and enabled accurate PIV measurements. \\

Matching the RI of the working fluid with that of the resin was found to be crucial for minimizing optical distortion and ensuring accurate velocity measurements.  
The experimental measurements showed that flow behavior strongly depends on both the channel geometry and the applied flow rate. Straight channels exhibited fully developed laminar flow with velocity profiles consistent with analytical Hagen–Poiseuille predictions. 

In the diseased geometries, the velocity distributions exhibited localized acceleration and deceleration corresponding to narrowing and widening regions. Increasing the flow rate resulted in proportionally higher velocity gradients and wall shear stresses, while the flow remained laminar for all tested conditions. These observations confirm that the developed optical setup is capable of resolving local flow variations within 3D printed fluidic devices.\\
Together, these developments enhance the physiological relevance, quantitative accuracy, and predictive capability of the 3D printed microlfudic devices, advancing their application in the study of cerebrovascular hemodynamics and diseases. \\

Future direction for the development of microfluidic vascular models should focus on several key areas. Extending the experimental setup from planar to three-dimensional geometries and implementing stereo-PIV would allow more comprehensive visualization of complex flow structures. Physiological realism can be enhanced by reducing channel diameters and incorporating compliant wall materials to study wall deformation effects. Utilizing patient-specific vascular geometries derived from medical imaging would further improve clinical relevance. Additionally, employing non-Newtonian fluids can better mimic blood rheology. Complementing experiments with computational fluid dynamics (CFD) simulations would provide deeper insights into flow patterns and shear stresses that are difficult to capture experimentally.

\section*{Author contributions}
\textbf{JvE}: Methodology, Software, Validation, Formal Analysis, Investigation, Resources, Data Curation, Writing - Original Draft, Visualization
\textbf{AS}: Methodology, Writing - Review \& Editing, Supervision
\textbf{DH}: Investigation
\textbf{SP}: Conceptualization, Writing - Review \& Editing, Supervision, Project Administration, Funding Acquisition
\textbf{PF}: Conceptualization, Writing - Review \& Editing, Visualization, Supervision, Project Administration, Funding Acquisition

\section*{Conflicts of interest}
There are no conflicts to declare.

\section*{Data availability}

The data that support the findings of this study are available from the corresponding author upon reasonable request.

\section*{Acknowledgements}

This work was supported by the Cohesion project (TUDelft). We sincerely thank Edwin Overmars and Christian Poelma for providing laboratory access, as well as for their valuable support with PIV data acquisition and interpretation.



\balance


\bibliography{rsc.bib} 
\bibliographystyle{rsc} 

\end{document}